\documentclass{PoS}

\usepackage{siunitx}%ang

\title{EUSO-TA ground based fluorescence detector: analysis of the detected events}

\ShortTitle{EUSO-TA: analysis of the detected events}

\author{\speaker{F.~Bisconti}$^1$, J.W.~Belz$^2$, M.E.~Bertaina$^{1,3}$, S.~Blin-Bondil$^4$, F.~Capel$^5$, M.~Casolino$^{6,7}$, T.~Ebisuzaki$^6$, J.~Eser$^8$, P.~Gorodetzky$^9$, J.N.~Matthews$^2$, E.~Parizot$^9$, L.W.~Piotrowski$^6$, Z.~Plebaniak$^{10}$, G.~Pr\'ev\^ot$^{9}$, M.~Putis$^{11}$, H.~Sagawa$^{12}$, N.~Sakaki$^6$, H.~Shin$^{12}$, K.~Shinozaki$^{1,3}$, P.~Sokolsky$^2$, Y.~Takizawa$^6$, Y.~Tameda$^{13}$, G.B.~Thomson$^2$,\newline for the JEM-EUSO Collaboration\footnote{for collaboration list see PoS(ICRC2019)1177}\\
$^1$National Institute for Nuclear Physics - Section of Turin, Turin, Italy; $^2$University of Utah, Salt Lake City, USA; $^3$University of Turin, Turin, Italy; $^4$Omega, Ecole Polytechnique, CNRS/IN2P3, Palaiseau, France; $^5$KTH Royal Institute of Technology, Stockholm, Sweden; $^6$RIKEN, Wako, Japan; $^7$National Institute for Nuclear Physics - Section of Rome Tor Vergata, Rome, Italy; $^8$Colorado School of Mines, Golden, USA; $^9$APC, Univ Paris Diderot, CNRS/IN2P3, CEA/Irfu, Obs de Paris, Sorbonne Paris; $^{10}$National Centre for Nuclear Research, Lodz, Poland; $^{11}$Institute of Experimental Physics, Kosice, Slovakia; $^{12}$Institute for Cosmic Ray Research, University of Tokyo, Kashiwa, Japan; $^{13}$Department of Engineering Science, Faculty of Engineering, Osaka\\
E-mail: \email{francesca.bisconti@to.infn.it}}	 

\abstract{EUSO-TA is a ground-based florescence detector built to validate the design of an ultra-high energy cosmic ray fluorescence detector to be operated in space. EUSO-TA detected the first air shower events with the technology developed within the JEM-EUSO program. It operates at the Telescope Array (TA) site in Utah, USA. With the external trigger provided by the Black Rock Mesa fluorescence detectors of Telescope Array (TA-FDs), EUSO-TA observed nine ultra-high energy cosmic ray events and several laser events from the Central Laser Facility of Telescope Array and portable lasers like the JEM-EUSO Global Light System prototype. The reconstruction parameters of the cosmic ray events which crossed the EUSO-TA field of view (both detected and not detected by EUSO-TA), were provided by the Telescope Array Collaboration. As the TA-FDs have a wider field of view than EUSO-TA ($\sim$30 times larger), they allow the cosmic ray energy reconstruction based on the observation of most of the extensive air-shower profiles, including the shower maximum, while EUSO-TA only observes a portion of the showers, usually far from the maximum. For this reason, the energy of the cosmic rays corresponding to the EUSO-TA signals appear lower than the actual ones. In this contribution, the analysis of the cosmic-ray events detected with EUSO-TA is discussed.}

\FullConference{36th International Cosmic Ray Conference -ICRC2019-\\
	July 24th - August 1st, 2019\\
	Madison, WI, U.S.A.}

\begin{document}
\setcounter{page}{2}

\section{Introduction}\label{sec:intro}
The JEM-EUSO program has the objective to build a fluorescence cosmic ray detector designed to observe Ultra-High Energy Cosmic Rays (UHECRs) from space, on-board the International Space Station (ISS). The UHECRs of interest have energy of $\sim$$10^{20}$~eV, as at this energy they are not strongly deflected by intergalactic and galactic magnetic fields, and therefore it would be possible to identify directly the sources for the first time. However, the observation of such cosmic rays is challenging because of their low flux of $\sim$1~particle/km$^2$ per millennium. The probability to detect UHECRs is strongly higher from space with respect to ground, since the observed area projected on ground would be up to $\sim$$10^5$~km$^2$, much larger than any possible ground-based experiment, covering areas of $\sim$$10^3$~km$^2$. In this context, UHECRs can be detected observing the UV fluorescence light emitted by Nitrogen molecules when Extensive Air Showers (EASs) induced by UHECRs cross the atmosphere, which can be considered as a huge calorimeter. 

\begin{figure}[h]%* puts figures on two columns
	\centering	
	\includegraphics[height=5.5cm]{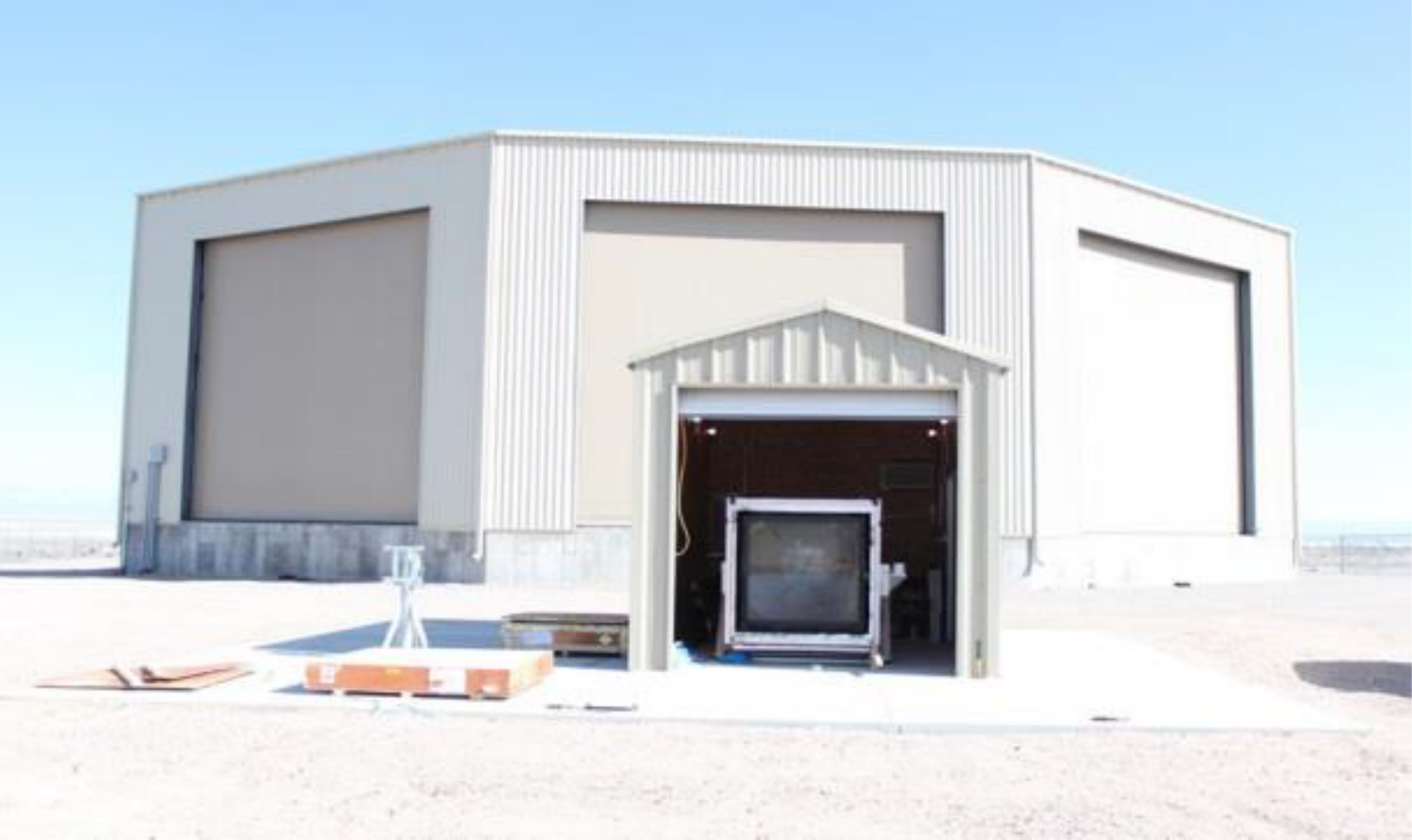}
	\caption{The EUSO-TA detector in front of the Black Rock Mesa Fluorescence Detector station.}
	\label{fig:eta} 
\end{figure}

EUSO-TA \cite{bib:euso-ta,bib:euso-ta_icrc2019} is one of the experiments of the JEM-EUSO program \cite{bib:jem-euso}, born and used to validate the observation principle and the design of this kind of detectors by observing EASs and laser pulses from ground. It is installed at the Telescope Array (TA) \cite{bib:ta} site in Utah (USA), in front of the Black Rock Mesa Fluorescence Detector (BRM-FD) station \cite{bib:ta-fd}, as shown in Figure~\ref{fig:eta}. With the external trigger provided by the BRM-FDs, it is possible to detect EASs and pulsed laser shots from the Central Laser Facility (CLF) \cite{bib:clf}, at about 20~km distance, which can be used to test and calibrate the detector. In addition, portable lasers with variable direction and energy, like the Global Light System prototype (GLS) \cite{bib:gls}, can be used to extend the range of distance of the source from the detector. The EASs detected by EUSO-TA have been analyzed to understand the performance and the detection limit of EUSO-TA. In this proceedings, updates with respect to the analysis discussed in reference \cite{bib:euso-ta} are given.

\section{The EUSO-TA detector}\label{sec:euso-ta_detector}
The EUSO-TA detector consists of an optical system with two flat Fresnel lenses of 1~m diameter and 8~mm thickness \cite{bib:lenses}. The EUSO-TA focal surface has a concave shape and consists of one Photo-Detector Module (PDM), which is a $\sim$17~cm$\,\times\,$17~cm active surface composed by an array of $6\times6$ Hamamatsu Multi-Anode Photo-Multiplier Tubes (MAPMTs, model R11265-M64) \cite{bib:mapmt}. Each MAPMT has $8\times8$~pixels of 2.88~mm side and a field of view of \ang{0.2}$\,\times\,$\ang{0.2}; the complete focal surface has 2304 pixels, and the total field of view of the detector is $\sim$\ang{10.6}$\,\times\,$\ang{10.6}. Each pixel has a gain (electron multiplication ratio) of more than $10^6$, which allows single photon counting and a photon detection efficiency of $\sim$30\%. A UV transmitting band pass filter in the range $290-430$~nm is glued on top of each MAPMT. 

Data are sampled in $2.5\,\mu\mbox{s}$ Gate Time Units (GTUs). The GTU corresponds to the time that light from the atmosphere takes to pass through a pixel of a detector in space at 400~km altitude, and represents the time resolution of the detectors within the JEM-EUSO program. The readout is performed by one 64-channel SPACIROC1 ASIC chip \cite{bib:spaciroc1} per MAPMT, with a dead time at beginning of each GTU of 0.2~$\mu$s and 30~ns double pulse resolution.

Typical of fluorescence detectors, EUSO-TA works at nighttime and in the best cases in clear sky conditions, to reduce effects of atmospheric and cloud attenuation. The elevation of the telescope axis can be manually changed from \ang{0} to \ang{30} with respect to the horizon, whereas its azimuth is fixed at \ang{53} from North counterclockwise, pointing to the CLF.

Data discussed in this proceedings have been acquired with the external trigger from  the BRM-FDs: in case of trigger, a packet of 128 GTUs centered around the time of the trigger is saved, which otherwise would be overwritten. This acquisition mode allows to know in advance when an EAS crossed the field of view of BRM-FDs, which encompasses the EUSO-TA one, but does not guarantee that the same event was observed also by EUSO-TA, as the field of view of BRM-FDs is about 30 times the EUSO-TA one. For this reason, offline analysis is required for the search of EAS events in the EUSO-TA data. Moreover, the TA analysis group performing shower reconstruction provide information as the arrival time and direction, impact point on ground, energy of the EAS events that cross the field of view of EUSO-TA, which can be used for further analyses.

The EUSO-TA experiment provides the opportunity to easily test the technology for existing and future experiments within the JEM-EUSO program, as it allows stable field observations for extended time periods.
To intensify the data acquisitions, an upgrade of the detector is foreseen in 2019. This will allow remote automated operations, making the collection of data possible continuously over the year. Furthermore, the sensitivity of the experiment will be enhanced, as the electronics will be improved with SPACIROC3 ASIC chips \cite{bib:spaciroc3}, with dead time at beginning of each GTU of 0.05~$\mu$s (instead of 0.2~$\mu$s for the currently used chip) and 5~ns double pulse resolution (instead of 30~ns). This will make the detector more efficient, in particular for the detection of close EASs, which cross individual pixels in short timescales. Indeed, if more photons hit the same pixel within the double pulse resolution, just the first hit is recorded: the result is that the event looks fainter than it could be with a shorter double pulse resolution. Moreover, the experiment will also be upgraded with advanced self-triggering capabilities \cite{bib:trigger}, replacing the current PDM data processing board with a new board with more memory and resources based on system-on chip (Zynq XC7Z030 FPGA \cite{bib:zynq}). It allows the implementation of data read-out on three timescales, to study different kind of phenomena. Data samples with time resolution of 2.5~$\mu$s will be still saved for UHECRs observations (the original time resolution of the JEM-EUSO detectors); data samples with resolutions of 320~$\mu$s (128~time frames of 2.5~$\mu$s) and 40.96~ms (128~time frames of 320~$\mu$s) are stored for slow events such as strangelets and meteors, identifiable offline with dedicated event search algorithms. 

\section{Observation of UHECR events}\label{sec:cr_det_sim}
Four data acquisition campaigns in the year 2015 and one in 2016 were done with the external trigger provided by BRM-FDs, for a total of about 140~hours. The analysis reported in this proceedings refers to data taken in 2015, equivalent to about 120~hours.
 
%Nine events have been detected. 
Energy and distance of the EASs crossing the field of view of EUSO-TA, whether detected or not, can be used to estimate the detection limit of the detector. As the field of view of BRM-FDs includes the EUSO-TA one, they detect all the EASs in the EUSO-TA field of view. The Telescope Array Collaboration provides the list of such events and also the reconstruction parameters of the UHECRs generating the EASs, reconstructed in monocular mode (using just data collected with the BRM-FDs). The reconstruction parameters used in this analysis are the impact parameter $R_{p}$, i.e. the shortest distance between the EAS axis and the detector in the shower-detector plane; the reconstructed energy of the primary particle $E_{recTA}$; the impact point on ground of the EAS axis; the zenith and azimuth angles of the EAS axis. 
\begin{figure}[h]%* puts figures on two columns
	\centering	
	\includegraphics[height=6.2cm]{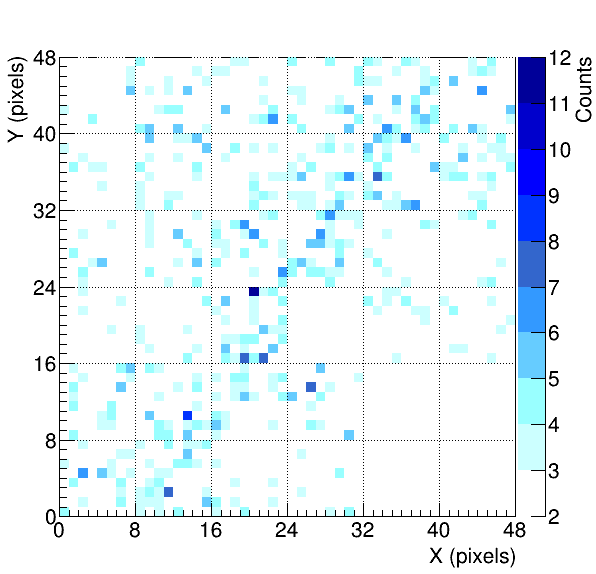}
	\includegraphics[width=8.2cm]{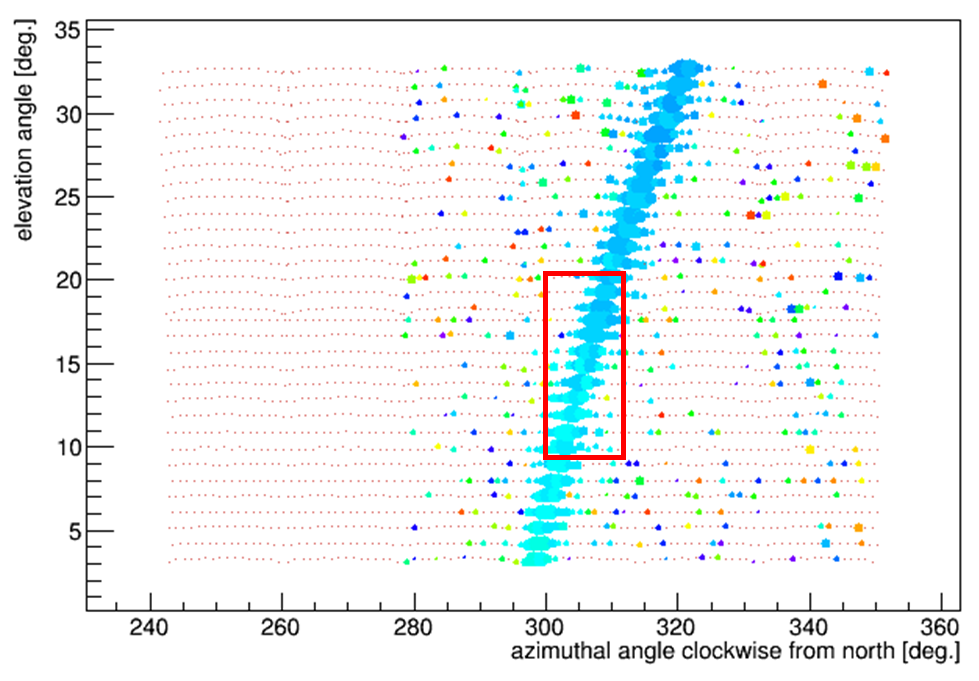}	
	\caption{The UHECR event detected on May 13th, 2015, with impact parameter $R_{p}=2.5$~km and energy $E_{recTA}=10^{18}$~eV. \textit{Left:} event detected by EUSO-TA, showing counts per pixel per GTU on the full PDM. \textit{Right:} event detected by BRM-FDs, in horizontal coordinates, where each circle represents one PMT of the BRM-FDs. the red rectangle indicates the EUSO-TA field of view.} \label{fig:ta_eta_event}
\end{figure}

The plot on the left in Figure~\ref{fig:ta_eta_event} shows a sample EAS detected by EUSO-TA in 1~GTU, with the counts per pixel per GTU on the full PDM. The plot on the right shows the same event detected by BRM-FDs, in horizontal coordinates, with the EUSO-TA field of view indicated by the red rectangle and each circle represents one PMT of the BRM-FDs. One can see that the spatial resolution is higher for EUSO-TA than for the BRM-FDs.
As the field of view of the BRM-FDs covers \ang{33} in elevation and \ang{100} in azimuth, the event reconstruction is based on the observation of the whole or most of the shower, including the shower maximum, i.e. the point along the shower longitudinal axis with the maximum number of particles. From the reconstruction of the EAS, it is possible to estimate the energy of the primary particle. On the other hand, EUSO-TA has a field of view of \ang{10.6}$\times$\ang{10.6} in elevation and azimuth, which is within the field of view of the BRM-FDs. This means that in case of an EAS in its field of view, EUSO-TA observes just a small portion of it, and in most cases it does not observe the maximum. This is not the case for space based experiments, which observe the entire EAS and for which the energy detection threshold is given by the number of photon counts coming from the maximum of the detected EAS. Instead, considering the photon counts detected by EUSO-TA, the energy of the primary particle would result in an equivalent energy $E_{eq}$ lower than the energy reconstructed by BRM-FDs $E_{recTA}$, due to the partial observation of the EAS. The equivalent energy $E_{eq}$ is the energy that an EAS would have based on its partial observation by EUSO-TA, as if the detector has the shower maximum in its field of view. To understand the detection limit of EUSO-TA, it is necessary to consider $E_{eq}$ instead of $E_{recTA}$. 
\begin{figure*}[p]%* puts figures on two columns
	\centering
	\includegraphics[width=12.2cm]{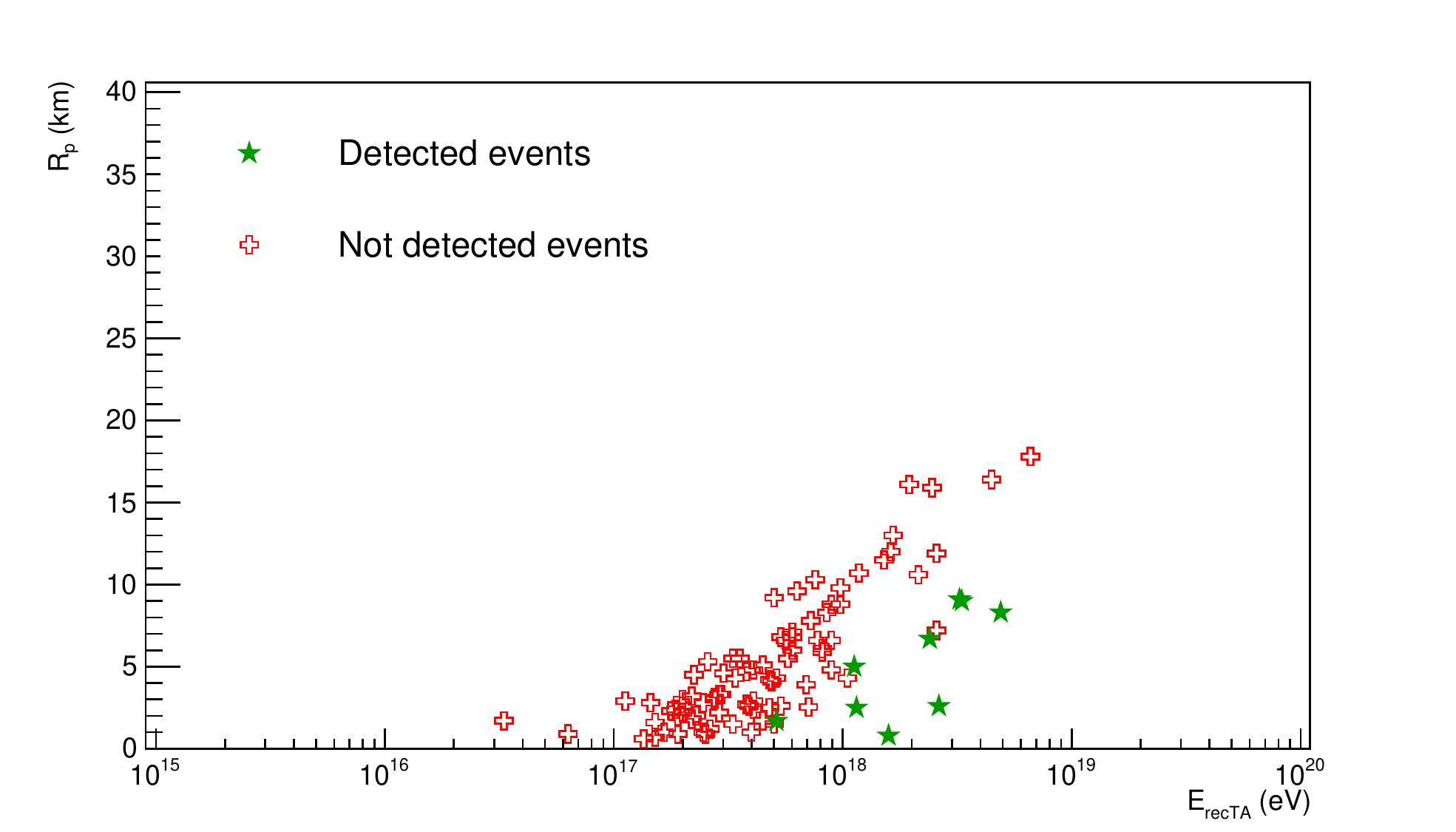}
	\includegraphics[width=12.2cm]{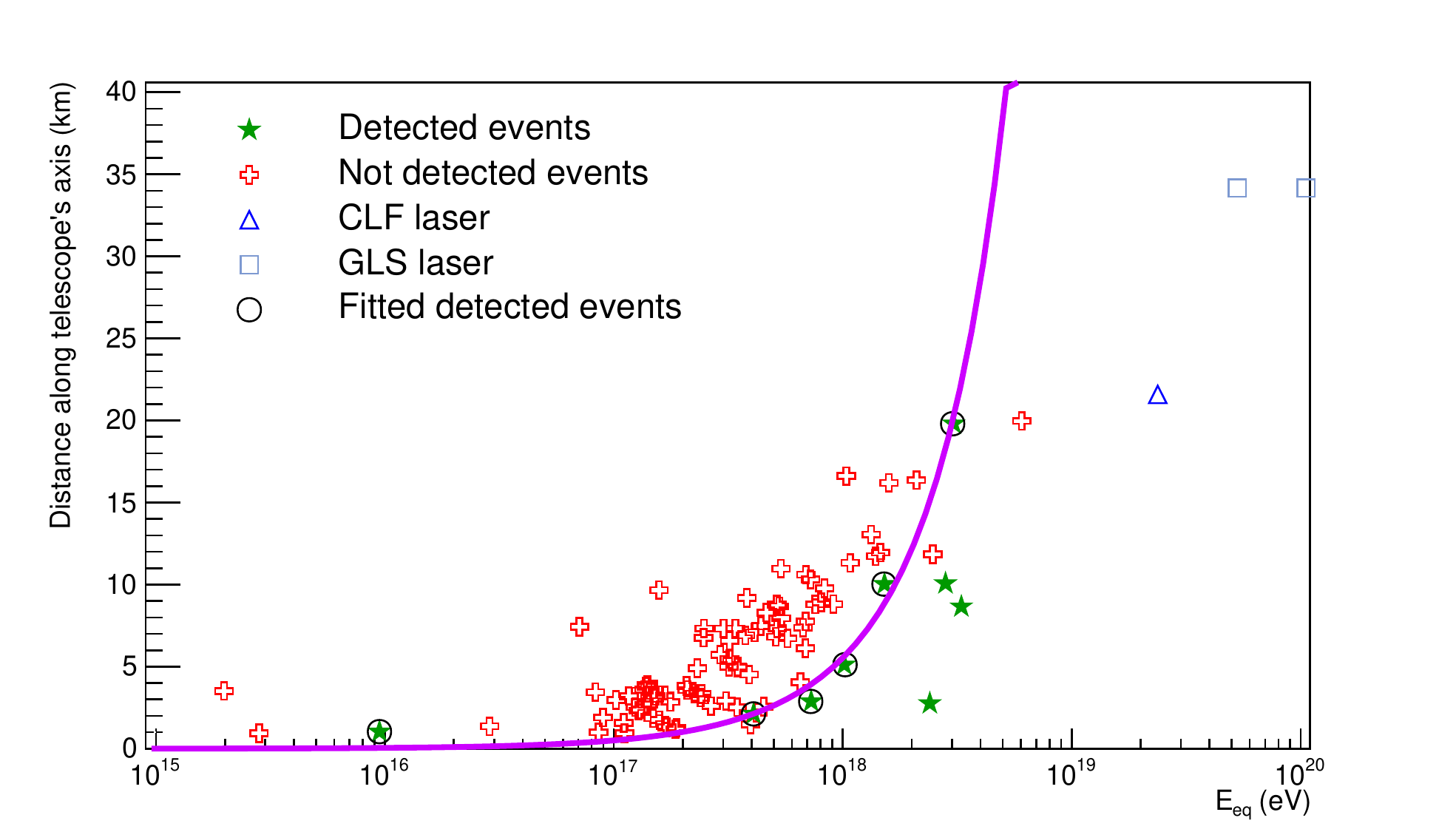}
	\includegraphics[width=12.2cm]{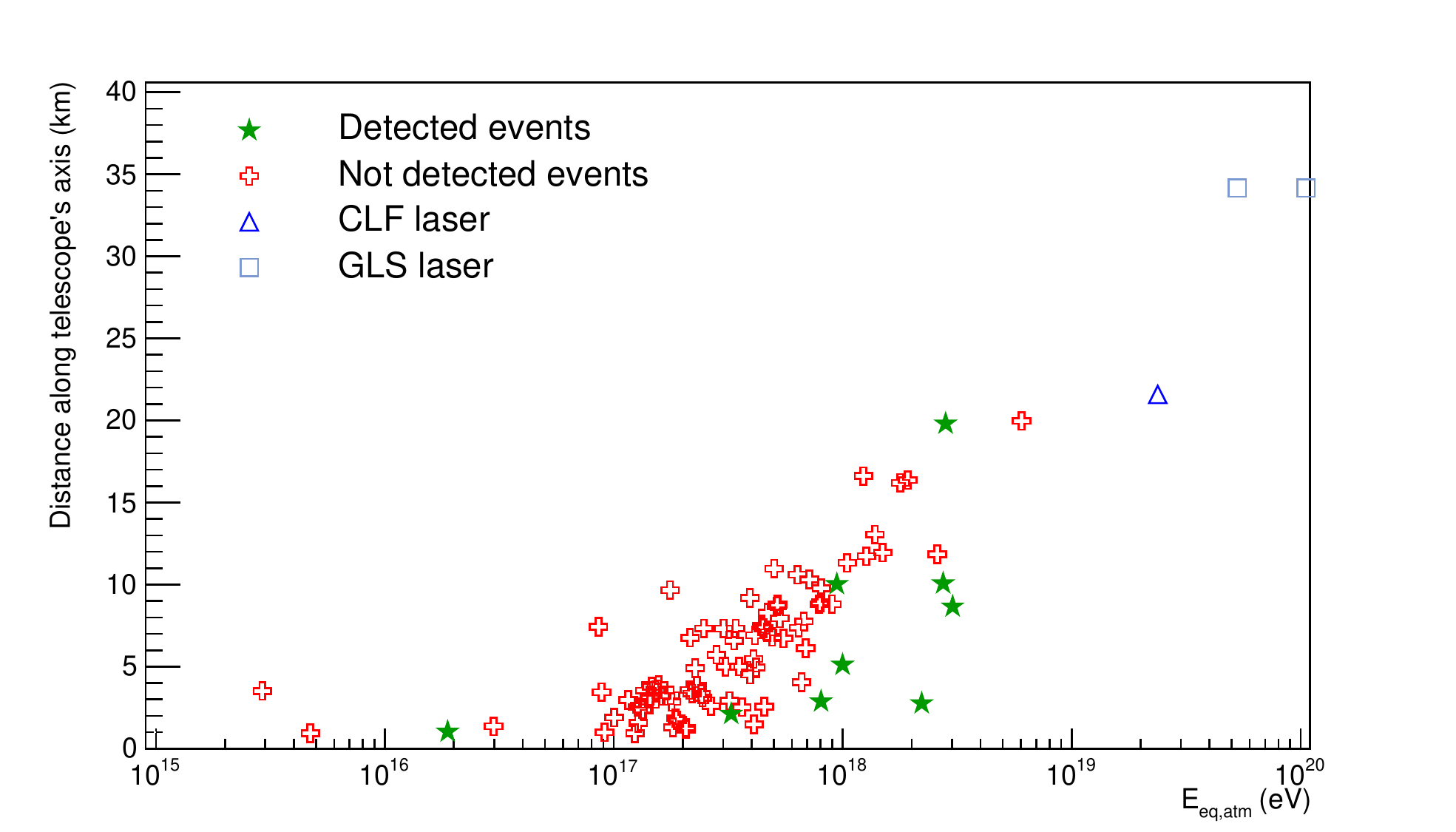}	
	\caption{EASs crossing the field of view of EUSO-TA. \textit{Top:} impact parameter versus reconstructed energy; \textit{Middle:} distance along the telescope axis versus equivalent energy; \textit{Bottom:} distance along the telescope axis versus equivalent energy corrected by the atmospheric attenuation.} \label{fig:limit}
\end{figure*}
\section{Analysis of the detection threshold}
In the previous analysis, the impact parameter $R_{p}$ was considered as distance of the EAS from the detector. The top panel of Figure~\ref{fig:limit} shows the EASs which crossed the EUSO-TA field of view with respect to their $E_{recTA}$ and $R_{p}$. Red crosses represent the non-detected events and green stars represent the detected events. A separation between detected and non-detected events is visible, as the detected events are with higher energy and lower impact parameter. In addition, laser events from the CLF and the GLS (with two energies), are plotted with triangles and squares, respectively. As the laser events are clearly visible above the night sky background, they stay well away on the right side of the line. However, the actual distance photons travel through along the telescope axis is more appropriate to find a relation between energy and distance. In particular for inclined showers approaching the detector, the distance along the telescope axis is much longer than the impact parameter.

To estimate $E_{eq}$ of the EASs that crossed the EUSO-TA field of view, an energy conversion factor is calculated and applied to $E_{recTA}$. Using the zenith, azimuth and energy of the EASs provided by the Telescope Array Collaboration, EAS simulations were performed with CONEX \cite{bib:conex}. These simulations provide the distribution of the energy loss (proportional to the number of photons emitted) with respect to the altitude. From the direction and impact point on ground of an EAS and the elevation angle of EUSO-TA during the data acquisition, event by event, it is possible to calculate the altitude at which EUSO-TA observed (or did not) such EAS. The energy conversion factor is the ratio between the energy loss of the EAS at the altitude where EUSO-TA observed with the actual elevation (calculated as the point where the distance between the telescope's optical axis and the shower axis is minimum), and the energy loss in the case the detector was pointing to the EAS maximum, with known altitude in the simulations. The middle panel of Figure~\ref{fig:limit} shows the same events, but plotted with their equivalent energy $E_{eq}$ and the distance along the telescope axis to the EASs. Here the separation between non-detected and detected events is more visible. As the equivalent energy of the shower is usually lower than the reconstructed energy, all points move to lower energies with respect to the former plot. Moreover, since the distance of the shower along the telescope axis is usually longer than the impact parameter, all the points move towards longer distances. An analytical fit of the upper detected events (highlighted with black circles) was made with a second order polynomial. All the detected events are supposed to be below and to the right of the line. A few non-detected events also appear in this region. This can be connected to the dead time of 0.2~$\mu$s the electronics has at the beginning of the GTUs, which makes it possible that the detector misses $\sim 10\%$ of the events, or on the uncertainties of the EAS reconstruction parameters.

The atmospheric absorption depending on the depth traversed by photons has to be considered, too. This takes into account that the distance along the EUSO-TA optical axis pointing to the EAS with the actual elevation angle at the time of the acquisition was different from the distance pointing to the shower maximum. Differences in the atmospheric depth reflects into differences in the atmospheric density that photons traverse. The Linsley parametrization used in the CORSIKA simulation software \cite{bib:corsika} is considered, to retrieve the atmospheric depths at given altitudes. In a similar way to the method used for the calculation of the equivalent energy, atmospheric correction factors are calculated as the ratio between the transmission of the atmosphere in the direction along the EUSO-TA optical axis and pointing to the EAS maximum. The atmospheric correction factor can be in principle equal to 1 whether for a certain event EUSO-TA observed the shower maximum, but in general it is greater or lower than 1, depending on the direction of the shower with respect to the detector. Such atmospheric correction factor is multiplied by the equivalent energy of the showers. The bottom panel of Figure~\ref{fig:limit} shows the equivalent energy corrected for the atmospheric transmission.

\section{Summary and Conclusion}\label{sec:summary}
EUSO-TA demonstrates the performance of a new detector technology for the observation of UHECRs, based on Fresnel lenses and MAPMTs. The detector has registered, using BRM-FD triggers, nine UHECRs events during its four observational campaigns in 2015. The response of the detector was tested using UV laser shots mimicking extensive air showers. The nine registered UHECRs give hints on the EUSO-TA sensitivity, taking into account the equivalent energy of EAS and the atmospheric extinction. 

Moreover, an optimization of the detector will be achieved through the hardware upgrades in the near future. This will allow to operate EUSO-TA remotely to intensify the acquisition campaigns and will make the detector self triggering, to operate autonomously from the BRM-FDs.\\

\section*{Acknowledgments}
	
\noindent The support received by the Telescope Array Collaboration is deeply acknowledged.\\
This work was partially supported by Basic Science Interdisciplinary Research Projects of RIKEN and JSPS KAKENHI Grant (JP17H02905, JP16H02426 and JP16H16737), by the Italian Ministry of Foreign Affairs and International Cooperation, by the Italian Space Agency through the ASI INFN agreement n. 2017-8-H.0 and contract n. 2016-1-U.0, by NASA award 11-APRA-0058 in the USA, by the Deutsches Zentrum f\"ur Luft- und Raumfahrt, by the French space agency CNES, the Helmholtz Alliance for Astroparticle Physics funded by the
Initiative and Networking Fund of the Helmholtz Association (Germany), by Slovak Academy of Sciences MVTS JEM-EUSO as well as VEGA grant agency project 2/0132/17, by National Science Centre in Poland grant (2015/19/N/ST9/03708 and 2017/27/B/ST9/02162), by Mexican funding agencies PAPIIT-UNAM, CONACyT and
the Mexican Space Agency (AEM). Russia is supported by ROSCOSMOS and the Russian Foundation for Basic Research Grant n. 16-29-13065. Sweden is funded by the Olle Engkvist Byggm\"astare Foundation.


\newpage
\begin{thebibliography}{99}
\bibitem{bib:euso-ta} G. Abdellaoui et al. (JEM-EUSO Coll.), \textit{EUSO-TA - first results from a ground-based EUSO telescope}, Astropart. Phys. \textbf{102}, 98-111 (2018)
\bibitem{bib:euso-ta_icrc2019} L.W. Piotrowski (JEM-EUSO Coll.), \textit{Results and status of the EUSO-TA detector}, PoS(ICRC2019)\textbf{388} (2019)
\bibitem{bib:jem-euso} J.H. Adams Jr. et al. (JEM-EUSO Coll.), \textit{JEM-EUSO observational technique and exposure}, Exp. Astron., \textbf{40}, 117-134 (2015)
\bibitem{bib:ta} M. Fukushima et al., \textit{Telescope Array project for extremely high energy cosmic rays}, Prog. Theor. Phys. Suppl., \textbf{151}, 206-210 (2003)
\bibitem{bib:ta-fd} H. Tokuno et al., \textit{New air fluorescence detectors employed in the Telescope Array experiment}, Nucl. Instrum. Meth.~A, \textbf{676}, 54-65 (2012)
\bibitem{bib:clf} Y. Takahashi et al., \textit{Central laser facility analysis at the Telescope Array experiment}, AIP Conference proceedings, \textbf{1367}, 157-160 (2011)
\bibitem{bib:gls} P. Hunt et al. (JEM-EUSO Coll.), \textit{The JEM-EUSO global light system laser station prototype}, PoS(ICRC2015)\textbf{626} (2016)
\bibitem{bib:lenses} Y. Hachisu et al. (JEM-EUSO Coll.), \textit{Manufacturing of the TA-EUSO and the EUSO-Balloon lenses}, proceedings of ICRC2013, ID\textbf{1040} (2013) 
%\bibitem{bib:mapmt} H. Prieto-Alfonso et al. (JEM-EUSO Coll.), "Multi anode photomultiplier tube reliability assessment for the JEM-EUSO space mission", arXiv:1501.05908 [physics.ins-det] (2015)
\bibitem{bib:mapmt} H. Prieto-Alfonso et al. (JEM-EUSO Coll.), \textit{Multi anode photomultiplier tube reliability assessment for the JEM-EUSO space mission}, Reliability Engineering and System Safety \textbf{133} 137-145 (2015) 
\bibitem{bib:spaciroc1} H. Miyamoto et al. (JEM-EUSO Coll.), \textit{Performance of the SPACIROC front-end ASIC for JEM-EUSO}, proceedings of ICRC2013, ID\textbf{1089} (2013)
\bibitem{bib:spaciroc3} S. Blin-Bondil, et al., \textit{SPACIROC3: A Front-End Readout ASIC for JEM-EUSO cosmic ray observatory}, PoS(TIPP2014)\textbf{172} (2014)
\bibitem{bib:trigger} M. Battisti et al. (JEM-EUSO Coll.), \textit{Trigger developments for the fluorescence detector of EUSO-TA and EUSO-SPB2}, PoS(ICRC2019)\textbf{426} (2019)
\bibitem{bib:zynq} A. Belov et al. (JEM-EUSO Coll.), \textit{The integration and testing of the Mini-EUSO multi-level trigger system}, Advances in Space Research \textbf{62} 2966-2976 (2017) %arXiv:1711.02376 [astro-ph.IM] (2018)
%\bibitem{bib:euso-ta_events} F. Bisconti et al. (JEM-EUSO Coll.), \textit{Simulation study of the detected and expected events for the EUSO-TA fluorescence detector}, PoS(ICRC2017)\textbf{463} (2017)
\bibitem{bib:conex} T. Bergmann et al., \textit{One-dimensional hybrid approach to extensive air shower simulation}, Astropart. Phys., \textbf{26} 420-432 (2007)
\bibitem{bib:corsika} D. Heck et al., \textit{CORSIKA: A Monte Carlo code to simulate extensive air showers},
Wissenschaftliche Berichte FZKA-6019 (1998)
%\bibitem{bib:spb_test} A.L. Cummings, \textit{Field Testing for EUSO-SPB: Logistics and First Results}, Master's thesis, Colorado School of Mines, Arthur Lakes Library (2017)
\end{thebibliography}
\end{document}